\documentclass[journal=jacsat,manuscript=article]{achemso}

\usepackage[version=3]{mhchem} 
\usepackage[normalem]{ulem}
 
\DeclareUnicodeCharacter{2212}{-}
\author{Rakesh Dhama}
\affiliation{Faculty of Engineering and Natural Science, Photonics, Tampere University, 33720 Tampere, Finland}
\author{Mohsin Habib}
\affiliation{Faculty of Engineering and Natural Science, Photonics, Tampere University, 33720 Tampere, Finland}
\author{Alireza R. Rashed}
\affiliation{Faculty of Engineering and Natural Science, Photonics, Tampere University, 33720 Tampere, Finland}
\author{Humeyra Caglayan}
\affiliation{Faculty of Engineering and Natural Science, Photonics, Tampere University, 33720 Tampere, Finland}
\email{humeyra.caglayan@tuni.fi}

\title[An \textsf{achemso} demo]
{Hot electron generation and relaxation dynamics in hyperbolic meta-antennas}

\begin{document}


\begin{abstract}
Conventional plasmonic nanoantennas enable scattering and absorption bands at the same wavelength region, making their utilization to full potential impossible for both features simultaneously. In this regard, hyperbolic meta-antennas (HMA) offer independent, tunable and separate scattering and absorption resonance bands at different spectral ranges. Here, we take advantage of these separated spectral bands in HMA to enhance and modify the hot electron-based phenomena. On the one hand, we observed the photoexcited emission of particle plasmons by hot carriers; on the other, we utilized ultrafast transient absorption spectroscopy to characterize the dynamics of hot electrons. In comparison to corresponding plasmonic nanodisk antennas (NDA), extended plasmon-modulated photoluminescence (PM-PL) spectrum in HMA is achieved, attributing to their particular scattering spectra. Furthermore, the absorption band at longer wavelengths in HMA enhances the excitation efficiency of plasmon-induced hot electrons (HEs) in the near-infrared region (NIR) with an extended lifetime. It broadens the utilization of the visible/NIR range that can not be observed in conventional NDAs. Our results further show that such meta-antennas can generate plasmon-induced hot carriers on demand and provide insight into the optimization and engineering of the hot-carrier-assisted photochemistry.
\end{abstract}

\section{Introduction}
Plasmonic nanoantennas are well-known to enable extreme light confinement and enhanced electromagnetic field at the nanoscale. Their intense light-focusing properties, scattering, and the tailored photon density of states may all act to modify and enhance the efficiency of spontaneous emission of emitters \cite{7,8,9}. However, plasmonic nanoantennas enhance scattering and absorption simultaneously at the same wavelength region, limiting these architectures to a specific application based on the scattering or absorption process. Recently, nanoantennas obtained by nanostructuring bulk hyperbolic metamaterials with alternating layers of metal and dielectric have emerged as a unique platform to tune scattering, absorption, and local-field confinement. In particular, such hyperbolic nanoantennas can excite super-radiant electric dipolar mode and sub-radiant magnetic dipolar mode, enabling the modification (separation) of the scattering (radiative) and absorption (non-radiative) spectrum in the same architecture \cite{maccaferri2019hyperbolic, Francesco, uriel}. 

Another attractive property of plasmonic structures is their ability to generate hot carriers: hot electrons (HE) and hot holes (HH) through nonradiative plasmon decay. 
Such hot carriers can be utilized in a variety of light-harvesting applications such as photochemical reactions \cite{choi2017engineering}, photodetection \cite{knight2011photodetection} and photocatalysis \cite{han2018tunable}. However, HEs generated in plasmonic nanostructures suffer ultra-short life, low yield, and short mean free path due to ultrafast electron-electron scattering \cite{wu2015efficient,ratchford2019plasmon}. These hinder the efficiency of plasmon-induced hot-electron transfer. On the other hand, the traditional inorganic oxide semiconductors such as TiO\textsubscript{2} and ZnO extensively used for photochemical reactions can only absorb in the ultraviolet region and only utilize a small part of the spectrum \cite{zhang2017surface}. Therefore, designing the particular hyperbolic meta-antennas (HMA) type structures based on plasmonic metal/semiconductor or dielectric layers can broaden the absorption spectra range and enhance the excitation efficiencies of HEs in visible and NIR regions.

Additionally, plasmonic nanostructures accelerate electron-electron scattering and lead to emission by hot carriers of particle plasmons (PPs). In other words, excited d-band holes recombine nonradiatively with sp electrons, leading to the emission of PPs. These plasmons subsequently radiate, giving rise to the photoluminescence (PL) \cite{dulkeith2004plasmon,rashed2020plasmon}.
Such particular PL from gold nanostructures is known as plasmon-modulated photoluminescence (PM-PL) and is correlated to their scattering spectra. PM-PL can be modulated by their plasmon resonances \cite{hu2012plasmon} and has characteristics of anti-photobleaching and anti-photoblinking unlike conventional fluorophores and is also thermally robust. Although these characteristics can open up potential applications, especially in the near-infrared (NIR) region \cite{he2008nonbleaching,tong2010bright,hirsch2003sershen,lu2012plasmonic}, due to the increased value of energy mismatch between excitation laser and plasmon resonances PM-PL spectra are obtained at the visible wavelengths which limits its applications. There have been several attempts to extend PM-PL in the NIR region by increasing the plasmonic nano-antennas size \cite{rashed2020plasmon}, changing the shapes of nanostructures \cite{hu2012plasmon}, using the roughened surfaces \cite{boyd1986photoinduced, wan2015plasmon}. However, the spectra of PM-PL have remained elusive in the NIR region through the conventional plasmonic nanoantennas to date.




This work investigates the PM-PL from designed and fabricated multilayer metal-dielectric hyperbolic meta-antennas (HMA) based on gold/silica stacking layers, enabling well-defined and separated scattering and absorption bands. While, to distinguish the effect of the multilayer in HMA, as a reference, conventional plasmonic (gold) nanodisk antennas (NDA) of thickness equivalent to that of total metal layers of an HMA are also realized. 
Furthermore, we studied the enhanced lifetime of energetic hot electrons excited by ultrafast photons at the interband transition in both types of structures. The effect of the separate absorption band in HMA on photogeneration and relaxation of  HEs is also systematically investigated. This work explores the true potential of HMA with spectrally separated scattering and absorption regions. It brings dual functions such as PM-PL at longer wavelengths and an extended lifetime of hot electrons on a common platform compared to conventional plasmonic NDA with overlapped scattering and absorption bands. Moreover, such functionalized structures also open up a strategy to excite HEs with a specific lifetime for all-optical control of photocatalysis. Therefore such rational heterostructures designed by plasmonic and adsorbate/dielectric layers can be developed as a very particular platform to prolong and control the lifetime of HEs for the efﬁcient utilization of plasmon-induced hot carriers.    

\section{Results and Discussion}

\begin{figure} [ht]
     \centering
     \includegraphics[width=0.95\textwidth]{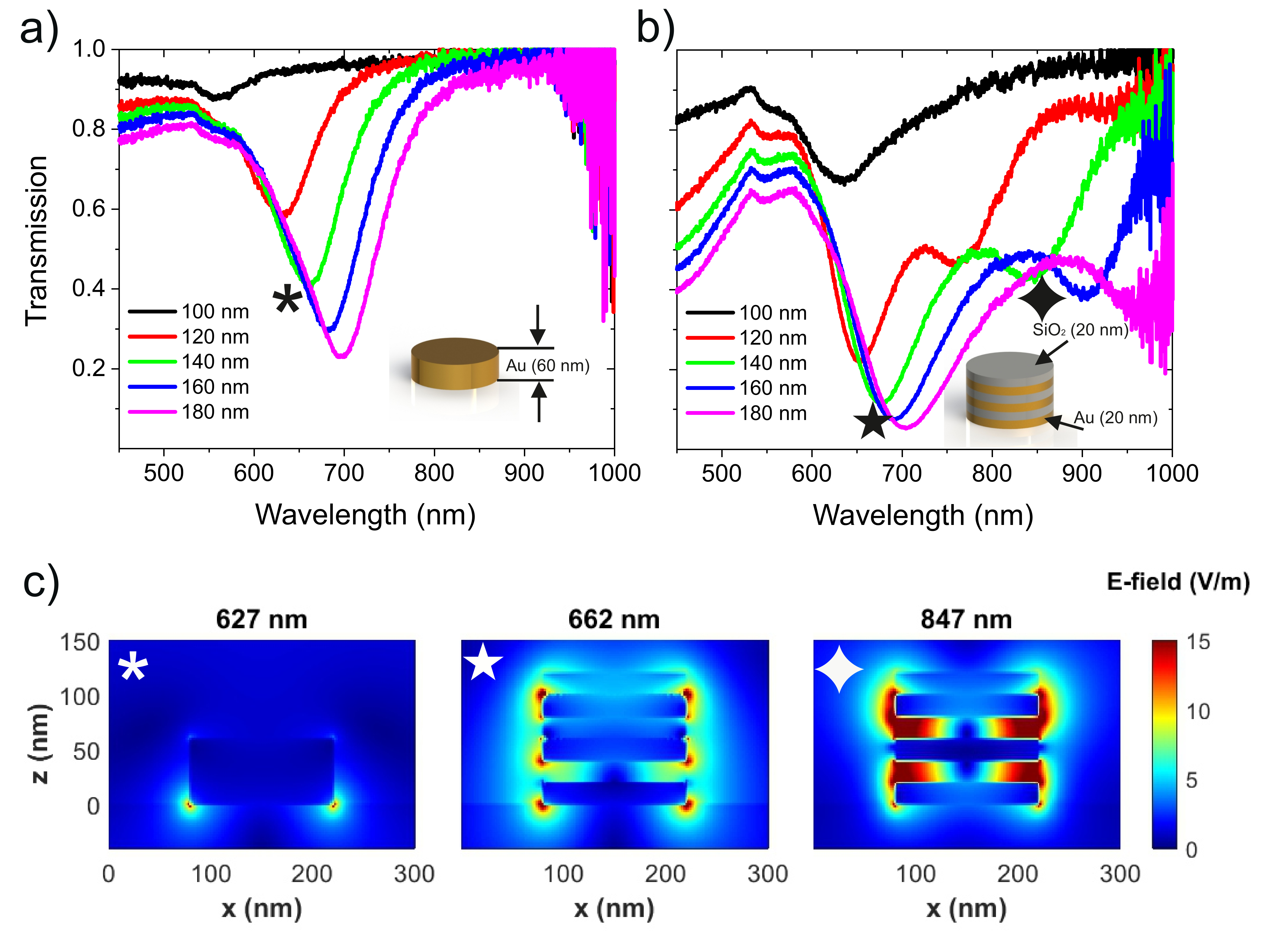}
     \caption{Measured transmission response of (a) NDAs, (b) HMAs with a diameter from 100 to 180 nm. (c) Electric field profiles in the \textit{xz} plane of NDA with 140 nm diameter at $\lambda$= 627 nm, 140 nm diameter HMA at $\lambda$= 662 nm, and 847 nm.}
     \label{fig:Fig1}
\end{figure}

To systematically investigate the features of the hot electrons in different plasmonic platforms, we designed and fabricated plasmonic nanodisks with Au thickness of 60 nm as well as meta-antennas based on multilayer metal-dielectric layers (3 bilayers of 20 nm Au and 20 nm SiO\textsubscript{2}). These multilayer hyperbolic antennas \cite{shekhar2014hyperbolic, Habib} are referred to as hyperbolic because of the intrinsic hyperbolic dispersion constituent deep subwavelength multilayers similar to type II hyperbolic metamaterials (See supplementary information). These
antennas do not show the spectral separation of radiative
and nonradiative channels without hyperbolic dispersion. 
Figure \ref{fig:Fig1} (a) and (b) present the measured transmission response of NDA and HMA for diameters from 100 to 180 nm with a step size of 20 nm, respectively. The transmission results show expectantly that the resonance in the transmission is red-shifted similarly for NDA and HMA as the diameter increases. However, the transmission results of HMAs differ as two resonances are observed, one in the similar range to the NDA resonance while the other is in the longer wavelengths. Furthermore, Figure \ref{fig:Fig1} (c), present cross-section field profiles in the \textit{xz} plane of NDA at $\lambda$= 627 nm, HMA at $\lambda$= 662 nm and 847 nm. 
These field profiles reveal that the first mode of HMA is similar to NDA with strong field confinement on the metallic layers, whereas the second mode is much more confined inside the dielectric layers.


Plasmonic nanoantennas provide a strong scattering attributed to the dipole electric mode. These structures are especially effective in exploiting the scattering while having a low parasitic absorption and tuning the resonance frequency to the desired wavelength range. We performed finite-difference-time-domain (FDTD) simulations to calculate the scattering and absorption for five different diameters (100, 120, 140, 160, and 180 nm) of HMA, similar to NDA (see Methods for details). 



\begin{figure} 
     \centering
     \includegraphics[width=1\textwidth]{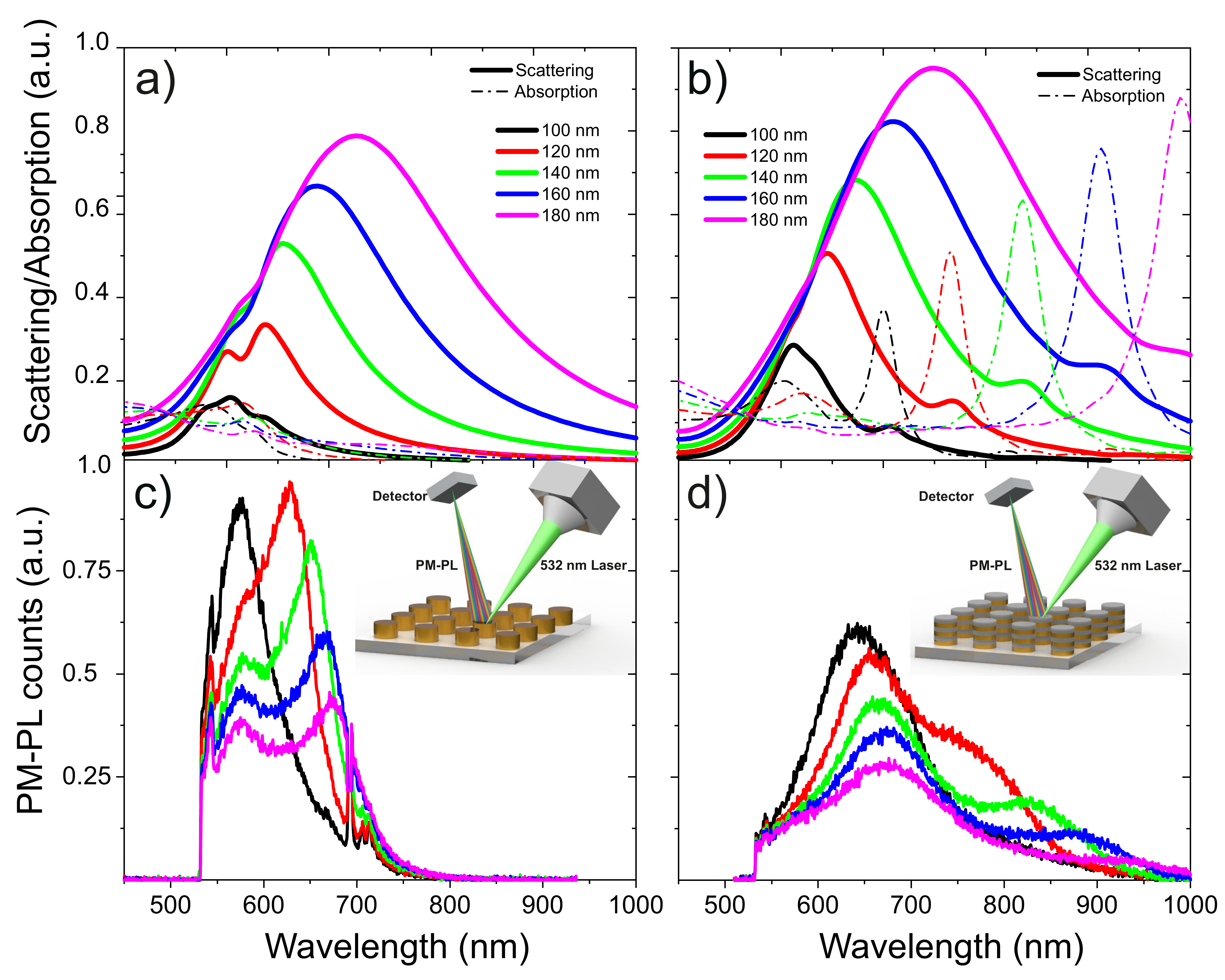}
     \caption{The calculated (a) scattering and absorption spectra of different sizes of NDA and (b) HMA structures. PM-PL spectra of (c) NDA structures (d) HMAs with diameters from 100 to 180 nm excited by a linearly polarized green laser ($\lambda$ = 532 nm).}
     \label{fig:Fig2}
\end{figure}

As seen in Figure \ref{fig:Fig2}, scattering increases as the nanoantenna size is enhanced and red-shifted, whereas the scattering of the HMA is higher than NDA. This trend has been commonly observed in experiments and reflects the direct dependence of the nanostructure extinction on the predominant dipole mode. At the same time, the absorption of NDA is almost identical from 100 to 180 nm diameters, while the absorption intensity and spectral position of HMAs are significantly changed with increasing nanoantenna diameter. The contribution of scattering to the total extinction (Scattering/Absorption) increases as absorption decreases for wavelengths more than 650 nm. The increase in the ratio of scattering to absorption with the nanostructure diameter is related to increased radiative damping in larger nanoparticles \cite{sonnichsen2002plasmon,scaffardi2004sizing,sonnichsen2002drastic}. These trends in plasmonic nanoantennas suggest that larger structures would be more suitable for applications based on light scattering. However, one needs to modify the absorption spectrum when applications are based on absorption contrast. In this regard, HMAs can provide a potential solution for reshaping the absorption and scattering properties. The inclusion of 20 nm SiO\textsubscript{2} in between 20 nm slices of Au enables the modification of scattering and absorption and scattering as shown in Figure \ref{fig:Fig2} (b). 

 
\subsection{Enhancement of Plasmon-Modulated Photoluminescence}
 

Once the optical properties of both plasmonic systems reveal that it is possible to modify the spectra using HMAs, we performed PM-PL measurements; as such, photoluminescence strictly follows the trends in the scattering of plasmonic structures, and its energy mismatch  between excitation laser and plasmon resonances peak ($\Delta$E = E\textsubscript{excitation} - E\textsubscript{plasmon})\cite{hu2012plasmon,rashed2020plasmon,ngoc2015plasmon}. An increase in $\Delta$E enables a significant decrease in PM-PL, which limits the fluorescence of metal nanostructures to the visible region. Figure \ref{fig:Fig2} (c) presents PM-PL measurements on NDA structures by exciting these structures using a linearly polarized green laser ($\lambda$ = 532 nm) and reports the decrease in PM-PL as the diameter of nanodisks increases. An increase in the energy mismatch of the excitation wavelength and the plasmon resonance leads to a significant decrease in PM-PL. Therefore, for the smaller NDA structure, non-equilibrium electrons are highly populated to excite particle plasmons with the minimal value of energy mismatch and lead to the intensification of PM-PL at a shorter wavelength. On the other hand, Figure \ref{fig:Fig2} (c) further shows weaker photoluminescence spectra peaked at a longer wavelength and PM-PL approaches to minimum value within the visible region for a larger NDA structure. When the PM-PL measurements were performed on the HMA system under the same experimental conditions, broader PL spectra approached the NIR spectral region, which can be attributed to the reradiation of the scattering mode of the HMA structure. As presented in Figure \ref{fig:Fig2} (d), interestingly, such PL enhancement in NIR I (650-950nm) region cannot be achieved in NDA due to the absence of such peak (shoulder) in scattering spectra and an increase in energy mismatch. PM-PL comparison for 120 nm diameter of both antennas clearly indicates the broadening of PL spectra up to 150 nm (towards near NIR region) in HMA with respect to NDA.  

Hyperbolic meta structures with two independent and tunable scattering and absorption bands have great potential to offer multiple applications on a single platform. As the NIR wavelength region provides the maximal penetration of light through biological tissues, photoluminescence in the NIR region is crucial to developing anti-photobleached fluorescent probes for imaging. At the same time, the hyperbolic nature of HMA induces strong confinement of the electric field in the absorption band and enhances the absorption efficiency, which enhances the photo-thermal therapeutic capabilities.

\subsection{Modification in Ultrafast Dynamics of Hot Electrons}

Under the illumination of light on plasmonic systems, the coherent electron oscillation nonradiatively dephases and generates hot electrons on a time scale ranging from 1 - 100 fs. Hot electrons are generally those electrons that are not in thermal equilibrium with their immediate environment. It rapidly thermalizes to a Fermi−Dirac distribution via different time scales such as  electron-electron scattering  and electron−phonon scatterings (100 fs - 10 ps) that result in a higher lattice temperature followed by the slow  dissipation of heat to the environment (100 ps - 10 ns) \cite{brongersma2015plasmon}. Hot electrons have been utilized to trigger several  chemical and physical phenomena. Still, their novel applications are limited due to fast relaxation processes and low transfer efficiency from metals to acceptors. Thus, manipulating/engineering the spatial and temporal dynamics of HEs by particular plasmonic structures is key to developing exciting plasmon-induced hot carrier-based devices. 

To understand the spatial and temporal dynamics of photogenerated  hot electrons in NDA and HMA systems, ultrafast transient absorption (TA) spectroscopic pump-probe setup equipped  with Astrella Ti: Sapphire laser (pulse duration 100 fs, repetition rate 1kHz) and an optical parametric amplifier (OPA)  has been employed in transmission mode. For the excitation of interband transitions in NDA and HMA, 100 fs pump pulses of 400 nm wavelength from OPA excite the samples. Then, the temporally delayed visible probe beam (420 - 800 nm) is varied to measure a differential transient absorption spectrum ($\Delta$A) by comparing the extinction spectrum with and without the pump pulse, which is chopped at half the repetition rate. In addition, ultrafast temporal dynamics of the systems at the specific time delay between pump and probe pulses are also extracted.


\begin{figure} 
     \centering
     \includegraphics[width=1\textwidth]{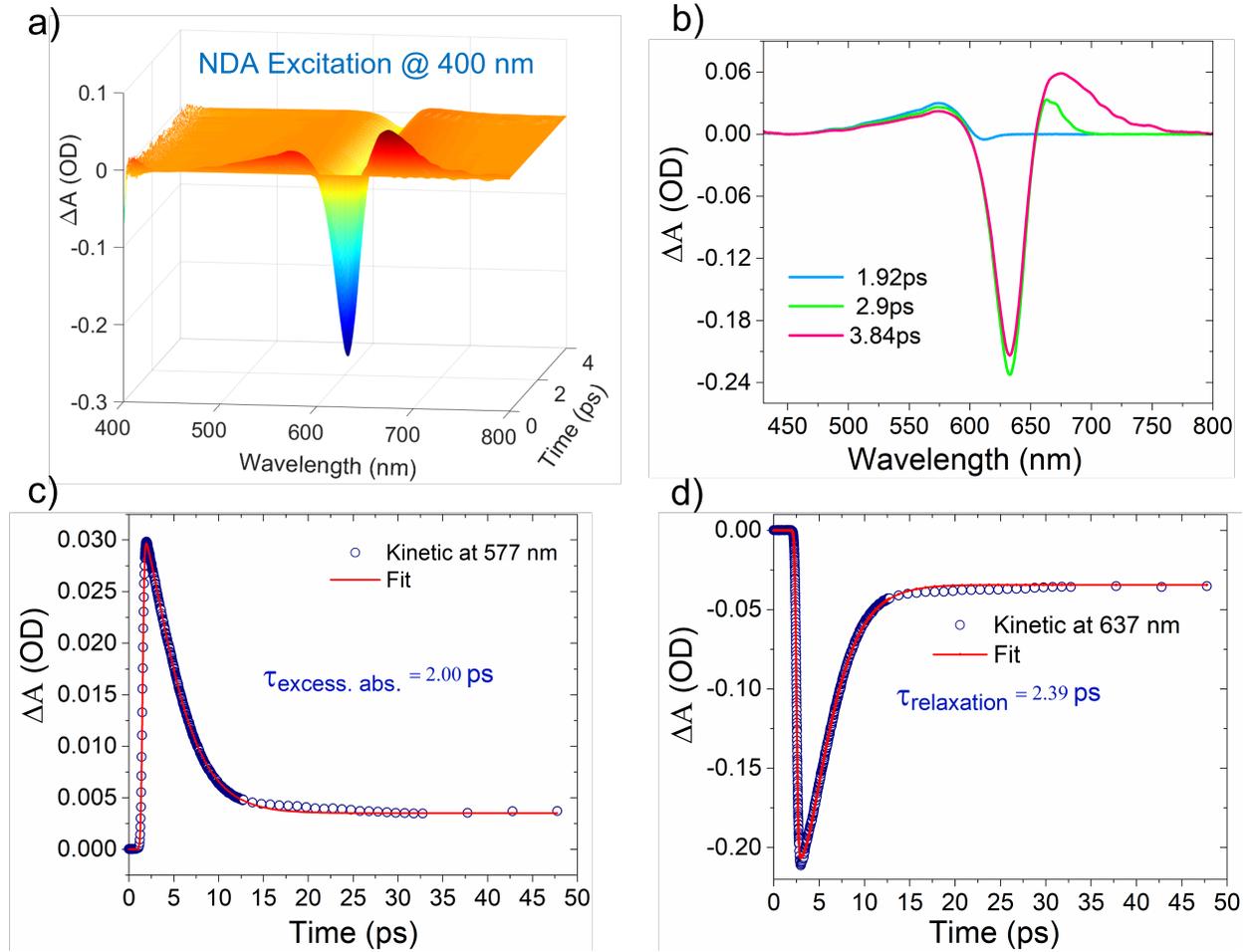}
     \caption{ Transient absorption response of nanodisk antenna system at interband excitation (a) A 3D surface panel for $\Delta$A spectra in different delay times for NDA system at the excitation of 400 nm wavelength pump pulses. (b) $\Delta$A spectra curves at specific time delays corresponding to the maximum  absorption values. (c, d) Temporal dynamics of the excessive absorption band at 577 nm and bleach region at 637 nm for the NDA system.}
     \label{fig:fig3}
\end{figure}
Figure \ref{fig:fig3} shows the spatial and temporal dynamics of HEs due to the interband transition in the plasmonic NDA system at the excitation of 400 nm (3.1 eV) pump pulses. Such excitation above the threshold energy (2.38 eV) for interband transition in gold \cite{schoenlein1987femtosecond} induces the electronic transition from 5d band to  hybridized 6sp band, resulting in a transient electron population in the conduction band. In this context, a  3D surface panel represents a bird's eye view of amplitude, wavelength, and time and is composed of  several transient absorption spectra recorded at a succession of closely-spaced time delays as presented in Figure \ref{fig:fig3} (a). Figure \ref{fig:fig3} (b) reports TA spectra curves in 3 different time delays (1.92 ps, 2.9 ps and 3.84 ps). TA spectra exhibit negative absorption ($\Delta$A) (bleach region) centered at the plasmon band of NDA, along with two positive absorption (excessive absorption) bands at lower and higher energy with respect to the  bleach region as shown in Figure \ref{fig:fig3} (b). The positive transient absorption band around 550 nm is attributed to interband excitation or thermal redistribution of thermal electrons below the Fermi level \cite{wang2019ultrafast}. The negative absorption band (bleach region) peaked around 640 nm, corresponding to the transition of electrons from lower energy levels to empty high energy states above Fermi level, while the positive transient spectral feature around 670 nm appears due to the absorption of hot electrons and generation of interband excitation induced plasmons \cite{logunov1997electron, zhang2018transient}. The time-resolved decay proﬁles of the excessive absorption band at 577 nm  and bleach region at 637 nm are extracted and fitted to almost similar times, 2.39 ps and 2.00 ps, respectively, as presented in Figure \ref{fig:fig3} (c) and (d), respectively. This means that time decay profiles of interband excitation and bleach region are similar as they are both affected by the cooling of the hot electrons in the system \cite{wang2019ultrafast}.

 \begin{figure} 
     \centering
     \includegraphics[width=1\textwidth]{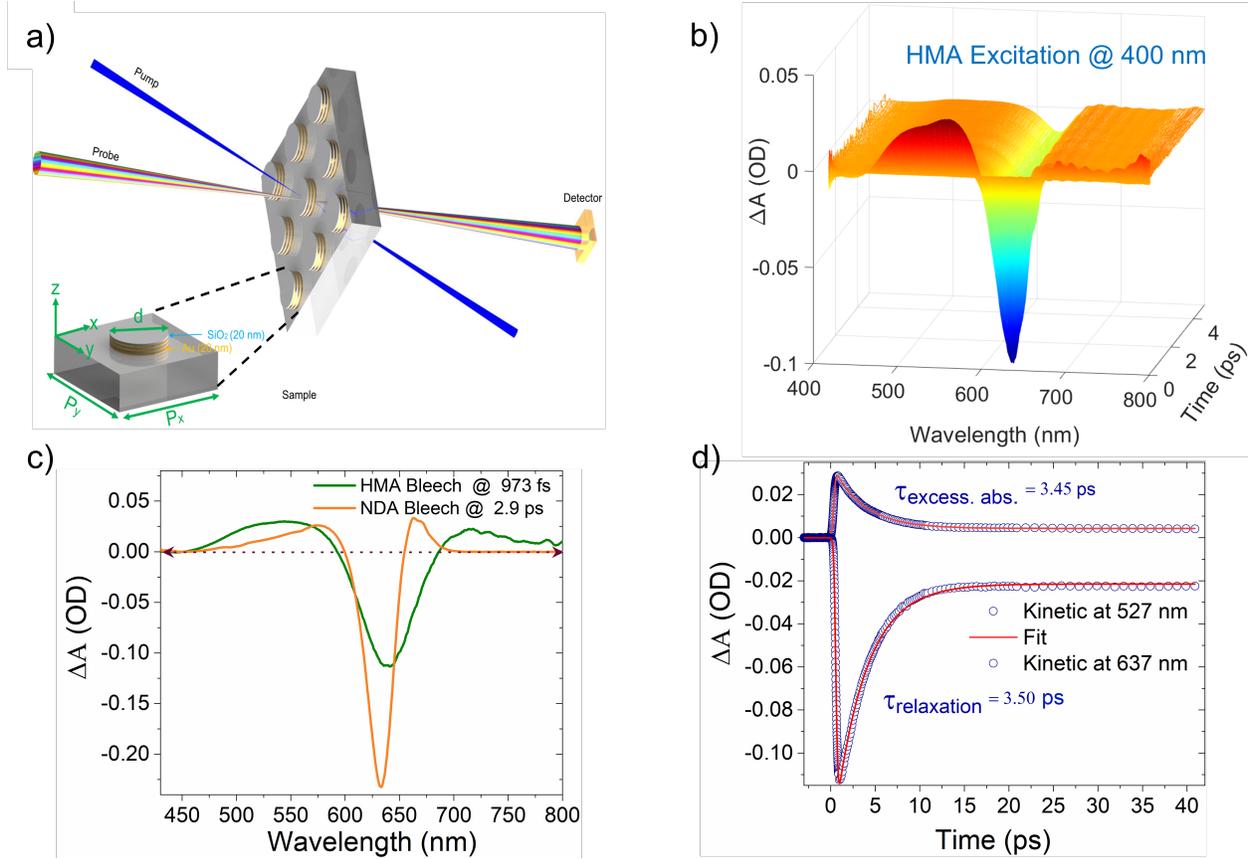}
     \caption{Transient absorption  response absorption  of hyperbolic meta antenna system at interband excitation (a) A 3D schematics of the HMA system at the excitation of 400 nm wavelength pump pulses and interrogated with delayed broadband probe beam. (b) A 3D surface panel for $\Delta$A spectra in different delay times for the HMA system at the excitation of 400 nm wavelength pump pulses. Extracted TA spectra curves of HMA AND NDA systems at their bleach region's maxima. (c, d) Temporal dynamics of the excessive absorption band at 527 nm and bleach region at 637 nm for the HMA system.}
     \label{fig:fig4}
\end{figure}
 Figure \ref{fig:fig4} (a) presents a schematic of experimental conﬁguration for HMA system when excited by ultrafast pump pulses (blue) and interrogated with time delayed broadband probe beam (rainbow) using the same pump power (400 nm excitation wavelength) and similar experimental conditions used to understand the TA response of the NDA system. A 3D TA spectra as a function of wavelength and time are shown in Figure \ref{fig:fig4} (b) and Figure \ref{fig:fig4} (c) reports TA spectra curves from  the bleach region's peak of NDA and HMA system to exhibit the broadening of transient response in HMA relative to NDA. Figure \ref{fig:fig4} (d) shows time-resolved decay proﬁles of the excessive absorption band at 527 nm  and bleach region at 637 nm, which are fitted to very similar times to 3.45 ps and 3.5 ps, respectively.

As followed in this work, in the comparison of the ultrafast response of NDA and HMA systems, one can clearly see a broad transient response in HMA (Figure \ref{fig:fig4} (c)) relative to the NDA system. This is attributed to the fact that an increase in the temperature enables red shift and broadening of LSPR in Au nanostructures \cite{yeshchenko2013temperature, wang2019ultrafast}. Thus, the thermalization of HEs by pump excitation enables the transient red-shift and broadening in the resonances of NDA and HMA. In particular, this phenomenon is more prominent in HMA due to the separated scattering and absorption band, enabling the hot electron generation and relaxation in the broad range with respect to NDA. It is also noteworthy that hot electron relaxation time (bleach region) in HMA is almost twice (3.5 ps) in comparison to that of NDA (2 ps) under the interband excitation (3.1 eV) at the same pump power.       

Hot electrons injection from metals into adsorbate/semiconductor have been extensively reported in Au/TiO\textsubscript{2} \cite{zhang2017surface, liu2013gold, brongersma2015plasmon, clavero2014plasmon}, Au/SiC\cite{hao2016visible} and Au/Si interfaces \cite{knight2011photodetection}. Due to the lower values of $\phi\textsubscript{SB}$ in these systems, a fraction of hot electrons cross the potential barrier to prolong their lifetime and trigger the chemical reactions in semiconductor/adsorbate before the recombination with holes in the metal. In contrast, the remaining HEs, that are unable to cross the potential barrier live relatively short and can not participate in photochemical transformations. Moreover, HEs significantly lose the energy to travel across the barrier, affecting photocatalysis efficiency on plasmonic/semiconductor heterostructures.

\begin{figure} 
     \centering
     \includegraphics[width=1\textwidth]{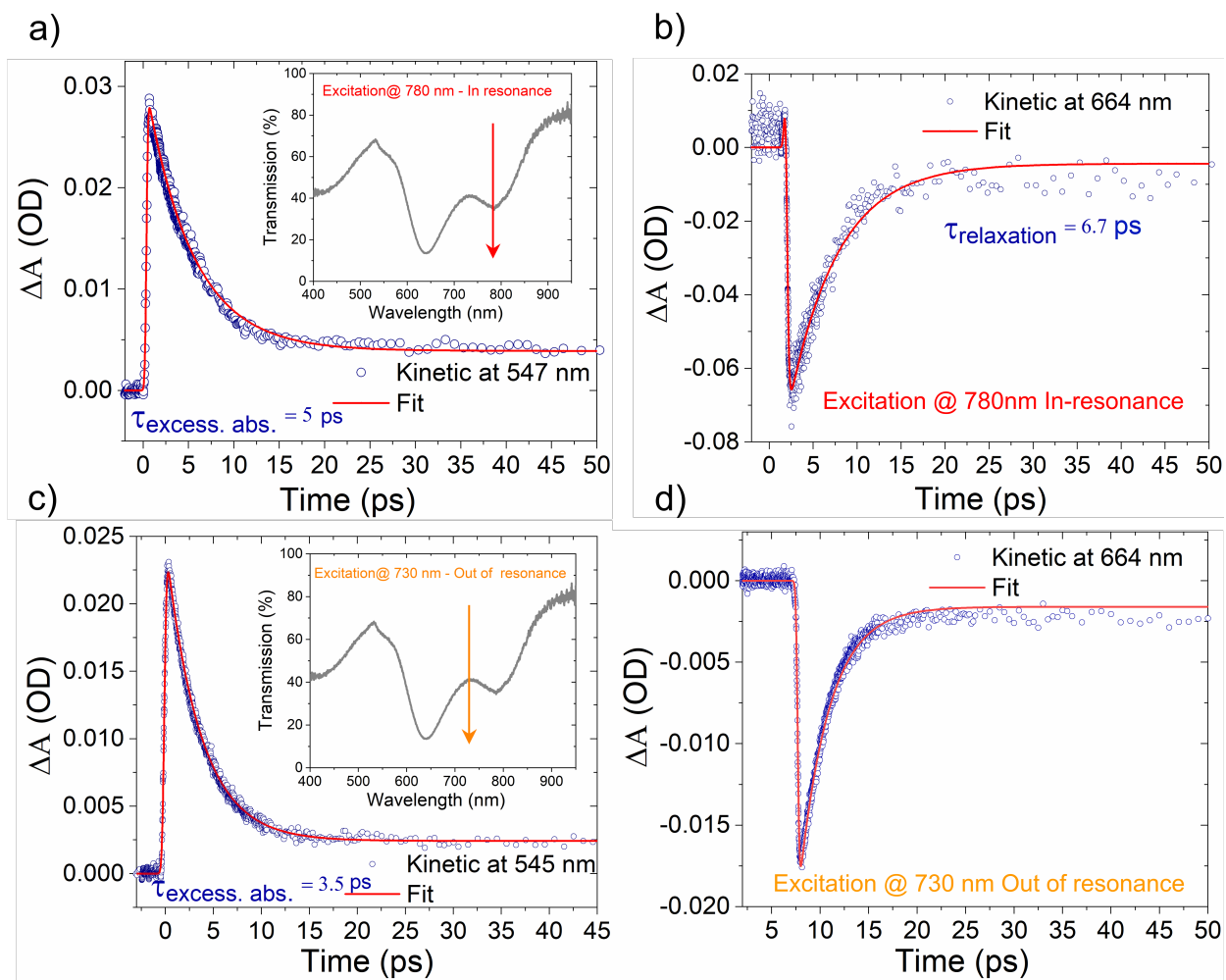}
     \caption{(a, b) Temporal dynamics of the excessive absorption band at 547 nm and bleach region at 664 nm for HMA system when excited by 780 nm wavelength pump pulses in the center of separate absorption band. (c, d) Temporal dynamics of the excessive absorption band at 545 nm  and bleach region at 664 nm for HMA system when excited by 730 nm wavelength pump pulses which are out of separate absorption band.}
     \label{fig:fig5}
\end{figure}

Thus, efficient hot electron generation and its accumulation for HE flux enhancement at the region of interest is a crucial issue that needs to be addressed for the improved performance and novel hot-electron-based mechanism. It has been shown that the rate of plasmon-induced H$\textsubscript{2}$ dissociation on Au NPs based on dielectric SiO$\textsubscript{2}$ \cite{mukherjee2014hot} is enhanced by two orders of magnitude higher than that observed on equivalently prepared AuNPs on TiO$\textsubscript{2}$ \cite{mukherjee2013hot}. This enhancement in dissociation eﬃciency of H$\textsubscript{2}$ is attributed to a large number of HEs present on the Au/SiO$\textsubscript{2}$ interface as HEs injection into wide bandgap dielectric SiO$\textsubscript{2}$ (9 eV) is not allowed and results in strong HE flux at the interface to enhance the chemical reactions in comparison to Au/TiO$\textsubscript{2}$ (Bandgap 3.1 eV). Such improvement in electron flux also leads to the elongation of the lifetime of HEs. 
In fact, the accumulation of excited hot electrons at the interfaces of gold/silica in HMA also explains the elongated HE's relaxation lifetime compared to that of NDA.
Additionally, HMA system advances the utilization of the spectrum due to the separate absorption band in NIR (see Fig. \ref{fig:Fig2} (c)) and excites the HEs with further elongated time at the excitation of the NIR pump pulses. Under similar circumstances, NDA does not exhibit any transient response due to the almost negligible absorption in the NIR region.

To investigate the effect of induced  absorption band on the temporal dynamics of HEs created in HMA, we have performed TA experiments on HMA (120 nm diameter) system excited by 780 nm wavelength (in-resonance) and 730 nm (out of resonance) pump pulses. Figure \ref{fig:fig5} (a) and (b) report time-resolved decay proﬁles of the excessive absorption band at 547 nm  and bleach region at 664 nm, fitted to time decay of 5 ps and 6.7 ps, respectively, when excited in so-called \textit{in-resonance band} (780 nm wavelength pump pulses) as clearly indicated in the inset of Figure \ref{fig:fig5} (a). Similarly, Figure \ref{fig:fig5} (c) and (d) show time-resolved decay proﬁles of excessive absorption band and bleach region at the excitation of \textit{out of resonance band} (730 nm wavelength pump pulses) as shown in the inset of Figure \ref{fig:fig5} (c). The fitted time decays are 3.5 ps and 3.0 ps, respectively, for the different observed wavelengths. The results clearly confirm that absorption efficiency and induced field confinement of the plasmonic structures are also crucial factors in extending the lifetime of HEs.

The enhanced absorption feature of HMA at a separated absorption band and strong induced field enables the elongated HE's lifetime at the in-resonance excitation. While HEs live shortly at out-of-resonance excitation (730 nm) due to lower absorption and weaker field. By changing the thickness and composition  of plasmonic and dielectric layers, HMA can also efficiently generate HEs from UV to deep NIR wavelength ranges. 


 \section{Conclusion and outlook}
In conclusion, HMAs based on gold/silica layers have been designed, fabricated, and compared with their counterpart of NDAs in terms of hot electron generation and relaxation dynamics. We have exploited the separate scattering and absorption bands in HMA for the two specific purposes and compared their characteristics with NDA systems at the same time. Based on the PM-PL spectra in HMA and NDA, the PM-PL spectra of HMA extend to longer NIR wavelengths due to the presence of an additional peak (shoulder) in HMA's scattering spectra. NDA does not exhibit this feature, and PL from gold nanodisks is limited to the visible range. Moreover, the ultrafast TA results reveal that the lifetime of HEs in HMA is elongated due to the enhanced HE flux at gold/silica interfaces compared to NDA at interband excitation. HMA also enables photo-induced HEs with elongated lifetime at NIR wavelengths due to the presence of a separate absorption band, which is limited in NDA. Such architectures like HMA successfully broaden the utilization of solar spectrum compared to NDA and can be designed and optimized to excite specific lifetime of HEs on demand. Therefore, HMA offers enhanced functions through a single platform (photo-excitation of electrons in the structure), exploiting scattering spectra to enable extended PMPL spectra and broadening the absorption spectra with elongated HE lifetime due to the spectrally separated absorption band. Overall our results reveal that it is possible to generate plasmon-induced hot carriers on demand by taking advantage of the flexible control of the spectra for further optimization and engineering of the hot-carrier-assisted photochemistry.

\section{Methods}
\subsection{Numerical simulations}
We use the commercially available software, Anys Lumerical FDTD Solutions, for 3D electromagnetic simulations of scattering, absorption and transmission, and field profiles of the samples. The refractive index values are taken from experimental data available in the literature to model Au \cite{johnson1972optical}, and  SiO\textsubscript{2} \cite{malitson1965interspecimen}. To simulate the scattering and absorption, the boundary conditions (BCs) are set to perfect match layers (PMLs) in all directions. We use a Total-Field Scattered-Field (TFSF) source of wavelengths 450-1000 nm.  The fields inside the source are the sum of  incident fields and  scattered fields, while only the scattered fields are visible outside the source. The analysis group calculates the absorption cross-section using the optical theorem. The uniform meshing of 2 nm was used to calculate scattering and absorption. 

For transmission calculations, the unit cell of size 440 nm is illuminated by a linearly \textit{x}-polarized plane wave source of wavelengths 450-1000 nm. The boundary conditions (BCs) are set to periodic in along \textit{x}, \textit{y} axes, and perfectly matched layers (PMLs) in the direction propagation z. Conformal meshing is used in simulations, while a finer mesh constraint of 2 nm is employed in the region enclosing the NDA and HMA to get a better resolution. Similarly, cross-sectional \textit{E\textsubscript{z}} and \textit{H\textsubscript{y}} fields are calculated at single wavelength values along different planes.

\subsection{Fabrication}
We fabricated the NDA using standard electron beam lithography (EBL), followed by the metal deposition and lift-off process. However, for HMA, we first deposited an alternative layer of Au/SiO\textsubscript{2}, then performed the EBL followed by the deposition  of nickel (Ni) as a mask. Next, the lift-off process and reactive ion etching (RIE) of the unwanted layer have been done. 
The 100x100 $\mu$m\textsuperscript{2} write field is used to create the NDAs with the area dose of 300 $\mu$C/cm\textsuperscript{2}. The exposed samples are developed for 60 sec in 1:3 Metylisobutylketon(MIBK): IPA solution and 30 sec in IPA to stop the development process. The developed samples are loaded in the electron beam evaporation deposition chamber to deposit 1.5 nm of Ti and 60 nm of Au. The S-1165 remover is used to lift off the unwanted Au.

For HMA, the cleaned glass samples are coated with alternative layers of Au and SiO\textsubscript{2} using an electron-beam evaporator. For better adhesion, 1 nm of Ti is used before each layer. Once the multilayers are ready, a nanodisk array is formed using similar e-beam lithography parameters to NDA. The developed samples in this case are coated with 15 nm Ni. The lift-off is done to remove unwanted metal from the sample and form Ni disks that serve as a mask. To transfer these pattering to multilayers and form HMA, RIE is performed using oxford instruments plasma technology machine. Argon and Fluoroform (CHF\textsubscript{3}) gases are used in the process. The flow of both gases was 25 sccm, radio frequency (RF) power 200 W, and pressure of 30 mTorr, and the etching time was set to 15 minutes. The etching time was first optimized for multilayers and Ni case and 15 min. 

\subsection{Optical characterization}
\subsubsection{Transmission Measurements}
Transmission spectra are measured using a microscope from WiTec (alpha300 R- Confocal Raman Imaging). The samples are excited with a broadband light source (Energetiq EQ-99XFC LDLS, spectrum 190 nm to 2100 nm). The optical pump beam is focused on the sample surface by using a 20x objective (Zeiss NA=0.4) in normal incidence. To detect the transmission spectra, a 50x objective (Zeiss NA=0.75) is placed at the back focal plane to collect transmitted light in a normal direction. The collected light is coupled to an optical fiber connected to a spectrometer (Ocean Optics Flame, detection range 400 nm - 900 nm). A neutral density filter of 2 (AR coated from Thorlabs) is placed in the beam path to reduce the beam spot size and then focused on the back focal plane. We first measure the transmission spectra from the glass substrate. Then, we measure the transmission spectrum of the meta-antenna arrays, which is normalized to the spectrum of glass. 

\subsubsection{PM-PL Measurements}
PM-PL measurements on HMA and NDA hybrid  structures  are performed on a multi-functional WITec alpha300C confocal Raman microscopy system. The photoluminescence spectra of the control and main samples are acquired utilizing a VIS-NIR Flame detector (350–1000 nm) provided by Ocean Optics with an integration time of 300 ms. The samples are excited with a 532 nm CW laser, and the PL signals are guided to the detector through a 50×Zeiss EC ‘Epiplan’ DIC objective (NA = 0.75, WD =1.0 mm). 

\subsubsection{Transient absorption spectroscopy measurements}

Ultrafast time-resolved pump-probe spectroscopy was performed using an amplified Ti: sapphire laser system equipped with an optical parametric amplifier (OPA). This system produced 100 fs pulses at 1.00 kHz with a center wavelength of 800 nm. Most of the output (90$\%$) was directed to the OPA to generate tunable pump pulses in the UV-Visible to near-infrared spectral regions to excite interband and plasmonic resonances, respectively. The remaining 10$\%$ of output power travels through a delay-line to enable  controlled time diﬀerence between pump and probe pulses and converts into a broadband probe beam to interrogate the sample at the normal incidence in transmission mode. At the same time, The chopper-modulated pump pulse is spectrally as well as temporally overlapped with probe beam on the sample. At the same time,  the detector is triggered to detect every probe pulse and calculate the  absorption spectrum. Repetition rates of the pump and probe beams turn out to be  500 Hz and 1 kHz, respectively. Therefore, the eﬀect of the pump beam will be observed only in one of the two consecutive probe beams.


\begin{acknowledgement}
We acknowledge the financial support of the European Research Council (Starting Grant project aQUARiUM; Agreement No. 802986), Academy of Finland Flagship Programme, (PREIN), (320165).  R.D. acknowledges the ﬁnancial support of the H2020 Research and Innovation Programme (Marie Skłodowska-Curie MULTIPLY Project; Agreement No. 713694).
\end{acknowledgement}

\bibliography{achemso}

\providecommand{\latin}[1]{#1}
\makeatletter
\providecommand{\doi}
  {\begingroup\let\do\@makeother\dospecials
  \catcode`\{=1 \catcode`\}=2 \doi@aux}
\providecommand{\doi@aux}[1]{\endgroup\texttt{#1}}
\makeatother
\providecommand*\mcitethebibliography{\thebibliography}
\csname @ifundefined\endcsname{endmcitethebibliography}
  {\let\endmcitethebibliography\endthebibliography}{}
\begin{mcitethebibliography}{41}
\providecommand*\natexlab[1]{#1}
\providecommand*\mciteSetBstSublistMode[1]{}
\providecommand*\mciteSetBstMaxWidthForm[2]{}
\providecommand*\mciteBstWouldAddEndPuncttrue
  {\def\EndOfBibitem{\unskip.}}
\providecommand*\mciteBstWouldAddEndPunctfalse
  {\let\EndOfBibitem\relax}
\providecommand*\mciteSetBstMidEndSepPunct[3]{}
\providecommand*\mciteSetBstSublistLabelBeginEnd[3]{}
\providecommand*\EndOfBibitem{}
\mciteSetBstSublistMode{f}
\mciteSetBstMaxWidthForm{subitem}{(\alph{mcitesubitemcount})}
\mciteSetBstSublistLabelBeginEnd
  {\mcitemaxwidthsubitemform\space}
  {\relax}
  {\relax}

\bibitem[Emam \latin{et~al.}(2020)Emam, Mansour, Mohamed, and Mohamed]{7}
Emam,~A.~N.; Mansour,~A.~S.; Mohamed,~M.~B.; Mohamed,~G.~G. \emph{Nanoscience
  in Medicine Vol. 1}; Springer, 2020; pp 459--488\relax
\mciteBstWouldAddEndPuncttrue
\mciteSetBstMidEndSepPunct{\mcitedefaultmidpunct}
{\mcitedefaultendpunct}{\mcitedefaultseppunct}\relax
\EndOfBibitem
\bibitem[Ghopry \latin{et~al.}(2019)Ghopry, Alamri, Goul, Cook, Sadeghi, Gutha,
  Sakidja, and Wu]{8}
Ghopry,~S.~A.; Alamri,~M.; Goul,~R.; Cook,~B.; Sadeghi,~S.; Gutha,~R.;
  Sakidja,~R.; Wu,~J.~Z. Au Nanoparticles/WS2 Nanodomes/Graphene van der Waals
  Heterostructure Substrates for Surface-Enhanced Raman Spectroscopy. \emph{ACS
  Applied Nano Materials} \textbf{2019}, \emph{11}, 33390--33398\relax
\mciteBstWouldAddEndPuncttrue
\mciteSetBstMidEndSepPunct{\mcitedefaultmidpunct}
{\mcitedefaultendpunct}{\mcitedefaultseppunct}\relax
\EndOfBibitem
\bibitem[Norville \latin{et~al.}(2020)Norville, Smith, and Dawson]{9}
Norville,~C.~A.; Smith,~K.~Z.; Dawson,~J.~M. Parametric optimization of visible
  wavelength gold lattice geometries for improved plasmon-enhanced fluorescence
  spectroscopy. \emph{Applied Optics} \textbf{2020}, \emph{59},
  2308--2318\relax
\mciteBstWouldAddEndPuncttrue
\mciteSetBstMidEndSepPunct{\mcitedefaultmidpunct}
{\mcitedefaultendpunct}{\mcitedefaultseppunct}\relax
\EndOfBibitem
\bibitem[Maccaferri \latin{et~al.}(2019)Maccaferri, Zhao, Isoniemi, Iarossi,
  Parracino, Strangi, and De~Angelis]{maccaferri2019hyperbolic}
Maccaferri,~N.; Zhao,~Y.; Isoniemi,~T.; Iarossi,~M.; Parracino,~A.;
  Strangi,~G.; De~Angelis,~F. Hyperbolic meta-antennas enable full control of
  scattering and absorption of light. \emph{Nano letters} \textbf{2019},
  \emph{19}, 1851--1859\relax
\mciteBstWouldAddEndPuncttrue
\mciteSetBstMidEndSepPunct{\mcitedefaultmidpunct}
{\mcitedefaultendpunct}{\mcitedefaultseppunct}\relax
\EndOfBibitem
\bibitem[Zhao \latin{et~al.}(2021)Zhao, Hubarevich, Iarossi, Borzda, Tantussi,
  Huang, and De~Angelis]{Francesco}
Zhao,~Y.; Hubarevich,~A.; Iarossi,~M.; Borzda,~T.; Tantussi,~F.; Huang,~J.-A.;
  De~Angelis,~F. Hyperbolic Nanoparticles on Substrate with Separate Optical
  Scattering and Absorption Resonances: A Dual Function Platform for SERS and
  Thermoplasmonics. \emph{Advanced Optical Materials} \textbf{2021}, \emph{9},
  2100888\relax
\mciteBstWouldAddEndPuncttrue
\mciteSetBstMidEndSepPunct{\mcitedefaultmidpunct}
{\mcitedefaultendpunct}{\mcitedefaultseppunct}\relax
\EndOfBibitem
\bibitem[Indukuri \latin{et~al.}(2020)Indukuri, Frydendahl, Bar-David,
  Mazurski, and Levy]{uriel}
Indukuri,~S. R.~K.; Frydendahl,~C.; Bar-David,~J.; Mazurski,~N.; Levy,~U. WS2
  Monolayers Coupled to Hyperbolic Metamaterial Nanoantennas: Broad
  Implications for Light–Matter-Interaction Applications. \emph{ACS Applied
  Nano Materials} \textbf{2020}, \emph{3}, 10226--10233\relax
\mciteBstWouldAddEndPuncttrue
\mciteSetBstMidEndSepPunct{\mcitedefaultmidpunct}
{\mcitedefaultendpunct}{\mcitedefaultseppunct}\relax
\EndOfBibitem
\bibitem[Choi \latin{et~al.}(2017)Choi, Jeong, Lee, Kang, Kim, Nam, and
  Song]{choi2017engineering}
Choi,~J.~Y.; Jeong,~D.; Lee,~S.~J.; Kang,~D.-g.; Kim,~S.~K.; Nam,~K.~M.;
  Song,~H. Engineering Reaction Kinetics by Tailoring the Metal Tips of
  Metal--Semiconductor Nanodumbbells. \emph{Nano letters} \textbf{2017},
  \emph{17}, 5688--5694\relax
\mciteBstWouldAddEndPuncttrue
\mciteSetBstMidEndSepPunct{\mcitedefaultmidpunct}
{\mcitedefaultendpunct}{\mcitedefaultseppunct}\relax
\EndOfBibitem
\bibitem[Knight \latin{et~al.}(2011)Knight, Sobhani, Nordlander, and
  Halas]{knight2011photodetection}
Knight,~M.~W.; Sobhani,~H.; Nordlander,~P.; Halas,~N.~J. Photodetection with
  active optical antennas. \emph{Science} \textbf{2011}, \emph{332},
  702--704\relax
\mciteBstWouldAddEndPuncttrue
\mciteSetBstMidEndSepPunct{\mcitedefaultmidpunct}
{\mcitedefaultendpunct}{\mcitedefaultseppunct}\relax
\EndOfBibitem
\bibitem[Han \latin{et~al.}(2018)Han, Li, Tang, and Xu]{han2018tunable}
Han,~C.; Li,~S.-H.; Tang,~Z.-R.; Xu,~Y.-J. Tunable plasmonic core--shell
  heterostructure design for broadband light driven catalysis. \emph{Chemical
  science} \textbf{2018}, \emph{9}, 8914--8922\relax
\mciteBstWouldAddEndPuncttrue
\mciteSetBstMidEndSepPunct{\mcitedefaultmidpunct}
{\mcitedefaultendpunct}{\mcitedefaultseppunct}\relax
\EndOfBibitem
\bibitem[Wu \latin{et~al.}(2015)Wu, Chen, McBride, and Lian]{wu2015efficient}
Wu,~K.; Chen,~J.; McBride,~J.~R.; Lian,~T. Efficient hot-electron transfer by a
  plasmon-induced interfacial charge-transfer transition. \emph{Science}
  \textbf{2015}, \emph{349}, 632--635\relax
\mciteBstWouldAddEndPuncttrue
\mciteSetBstMidEndSepPunct{\mcitedefaultmidpunct}
{\mcitedefaultendpunct}{\mcitedefaultseppunct}\relax
\EndOfBibitem
\bibitem[Ratchford(2019)]{ratchford2019plasmon}
Ratchford,~D.~C. Plasmon-induced charge transfer: Challenges and outlook.
  \emph{ACS nano} \textbf{2019}, \emph{13}, 13610--13614\relax
\mciteBstWouldAddEndPuncttrue
\mciteSetBstMidEndSepPunct{\mcitedefaultmidpunct}
{\mcitedefaultendpunct}{\mcitedefaultseppunct}\relax
\EndOfBibitem
\bibitem[Zhang \latin{et~al.}(2017)Zhang, He, Guo, Hu, Huang, Mulcahy, and
  Wei]{zhang2017surface}
Zhang,~Y.; He,~S.; Guo,~W.; Hu,~Y.; Huang,~J.; Mulcahy,~J.~R.; Wei,~W.~D.
  Surface-plasmon-driven hot electron photochemistry. \emph{Chemical reviews}
  \textbf{2017}, \emph{118}, 2927--2954\relax
\mciteBstWouldAddEndPuncttrue
\mciteSetBstMidEndSepPunct{\mcitedefaultmidpunct}
{\mcitedefaultendpunct}{\mcitedefaultseppunct}\relax
\EndOfBibitem
\bibitem[Dulkeith \latin{et~al.}(2004)Dulkeith, Niedereichholz, Klar, Feldmann,
  Von~Plessen, Gittins, Mayya, and Caruso]{dulkeith2004plasmon}
Dulkeith,~E.; Niedereichholz,~T.; Klar,~T.; Feldmann,~J.; Von~Plessen,~G.;
  Gittins,~D.; Mayya,~K.; Caruso,~F. Plasmon emission in photoexcited gold
  nanoparticles. \emph{Physical Review B} \textbf{2004}, \emph{70},
  205424\relax
\mciteBstWouldAddEndPuncttrue
\mciteSetBstMidEndSepPunct{\mcitedefaultmidpunct}
{\mcitedefaultendpunct}{\mcitedefaultseppunct}\relax
\EndOfBibitem
\bibitem[Rashed \latin{et~al.}(2020)Rashed, Habib, Das, Ozbay, and
  Caglayan]{rashed2020plasmon}
Rashed,~A.~R.; Habib,~M.; Das,~N.; Ozbay,~E.; Caglayan,~H. Plasmon-modulated
  photoluminescence enhancement in hybrid plasmonic nano-antennas. \emph{New
  Journal of Physics} \textbf{2020}, \emph{22}, 093033\relax
\mciteBstWouldAddEndPuncttrue
\mciteSetBstMidEndSepPunct{\mcitedefaultmidpunct}
{\mcitedefaultendpunct}{\mcitedefaultseppunct}\relax
\EndOfBibitem
\bibitem[Hu \latin{et~al.}(2012)Hu, Duan, Yang, and Shen]{hu2012plasmon}
Hu,~H.; Duan,~H.; Yang,~J.~K.; Shen,~Z.~X. Plasmon-modulated photoluminescence
  of individual gold nanostructures. \emph{Acs Nano} \textbf{2012}, \emph{6},
  10147--10155\relax
\mciteBstWouldAddEndPuncttrue
\mciteSetBstMidEndSepPunct{\mcitedefaultmidpunct}
{\mcitedefaultendpunct}{\mcitedefaultseppunct}\relax
\EndOfBibitem
\bibitem[He \latin{et~al.}(2008)He, Xie, and Ren]{he2008nonbleaching}
He,~H.; Xie,~C.; Ren,~J. Nonbleaching fluorescence of gold nanoparticles and
  its applications in cancer cell imaging. \emph{Analytical chemistry}
  \textbf{2008}, \emph{80}, 5951--5957\relax
\mciteBstWouldAddEndPuncttrue
\mciteSetBstMidEndSepPunct{\mcitedefaultmidpunct}
{\mcitedefaultendpunct}{\mcitedefaultseppunct}\relax
\EndOfBibitem
\bibitem[Tong \latin{et~al.}(2010)Tong, Cobley, Chen, Xia, and
  Cheng]{tong2010bright}
Tong,~L.; Cobley,~C.~M.; Chen,~J.; Xia,~Y.; Cheng,~J.-X. Bright Three-Photon
  Luminescence from Gold/Silver Alloyed Nanostructures for Bioimaging with
  Negligible Photothermal Toxicity. \emph{Angewandte Chemie International
  Edition} \textbf{2010}, \emph{49}, 3485--3488\relax
\mciteBstWouldAddEndPuncttrue
\mciteSetBstMidEndSepPunct{\mcitedefaultmidpunct}
{\mcitedefaultendpunct}{\mcitedefaultseppunct}\relax
\EndOfBibitem
\bibitem[Hirsch \latin{et~al.}(2003)Hirsch, Stafford, and
  Bankson]{hirsch2003sershen}
Hirsch,~L.; Stafford,~R.; Bankson,~J. Sershen, nanoscale environment, the
  device comprising: SR; Rivera, B.; Price, RE; Hazle, JD; Halas, NJ; a solid
  state material in contact with the nanoscale envi West. \emph{JL Proc. Natl.
  Acad. Sci. USA} \textbf{2003}, \emph{100}, 13549--13554\relax
\mciteBstWouldAddEndPuncttrue
\mciteSetBstMidEndSepPunct{\mcitedefaultmidpunct}
{\mcitedefaultendpunct}{\mcitedefaultseppunct}\relax
\EndOfBibitem
\bibitem[Lu \latin{et~al.}(2012)Lu, Hou, Zhang, Liu, Shen, Luo, and
  Gong]{lu2012plasmonic}
Lu,~G.; Hou,~L.; Zhang,~T.; Liu,~J.; Shen,~H.; Luo,~C.; Gong,~Q. Plasmonic
  sensing via photoluminescence of individual gold nanorod. \emph{The Journal
  of Physical Chemistry C} \textbf{2012}, \emph{116}, 25509--25516\relax
\mciteBstWouldAddEndPuncttrue
\mciteSetBstMidEndSepPunct{\mcitedefaultmidpunct}
{\mcitedefaultendpunct}{\mcitedefaultseppunct}\relax
\EndOfBibitem
\bibitem[Boyd \latin{et~al.}(1986)Boyd, Yu, and Shen]{boyd1986photoinduced}
Boyd,~G.; Yu,~Z.; Shen,~Y. Photoinduced luminescence from the noble metals and
  its enhancement on roughened surfaces. \emph{Physical Review B}
  \textbf{1986}, \emph{33}, 7923\relax
\mciteBstWouldAddEndPuncttrue
\mciteSetBstMidEndSepPunct{\mcitedefaultmidpunct}
{\mcitedefaultendpunct}{\mcitedefaultseppunct}\relax
\EndOfBibitem
\bibitem[Wan \latin{et~al.}(2015)Wan, Wang, Yin, Li, Hu, Li, Shen, and
  Nijhuis]{wan2015plasmon}
Wan,~A.; Wang,~T.; Yin,~T.; Li,~A.; Hu,~H.; Li,~S.; Shen,~Z.~X.; Nijhuis,~C.~A.
  Plasmon-modulated photoluminescence of single gold nanobeams. \emph{ACS
  photonics} \textbf{2015}, \emph{2}, 1348--1354\relax
\mciteBstWouldAddEndPuncttrue
\mciteSetBstMidEndSepPunct{\mcitedefaultmidpunct}
{\mcitedefaultendpunct}{\mcitedefaultseppunct}\relax
\EndOfBibitem
\bibitem[Shekhar \latin{et~al.}(2014)Shekhar, Atkinson, and
  Jacob]{shekhar2014hyperbolic}
Shekhar,~P.; Atkinson,~J.; Jacob,~Z. Hyperbolic metamaterials: fundamentals and
  applications. \emph{Nano convergence} \textbf{2014}, \emph{1}, 14\relax
\mciteBstWouldAddEndPuncttrue
\mciteSetBstMidEndSepPunct{\mcitedefaultmidpunct}
{\mcitedefaultendpunct}{\mcitedefaultseppunct}\relax
\EndOfBibitem
\bibitem[Habib \latin{et~al.}(2020)Habib, Briukhanova, Das, Yildiz, and
  Caglayan]{Habib}
Habib,~M.; Briukhanova,~D.; Das,~N.; Yildiz,~B.~C.; Caglayan,~H. Controlling
  the plasmon resonance via epsilon-near-zero multilayer metamaterials.
  \emph{Nanophotonics} \textbf{2020}, \emph{9}, 3637--3644\relax
\mciteBstWouldAddEndPuncttrue
\mciteSetBstMidEndSepPunct{\mcitedefaultmidpunct}
{\mcitedefaultendpunct}{\mcitedefaultseppunct}\relax
\EndOfBibitem
\bibitem[S{\"o}nnichsen \latin{et~al.}(2002)S{\"o}nnichsen, Franzl, Wilk,
  Von~Plessen, and Feldmann]{sonnichsen2002plasmon}
S{\"o}nnichsen,~C.; Franzl,~T.; Wilk,~T.; Von~Plessen,~G.; Feldmann,~J. Plasmon
  resonances in large noble-metal clusters. \emph{New Journal of Physics}
  \textbf{2002}, \emph{4}, 93\relax
\mciteBstWouldAddEndPuncttrue
\mciteSetBstMidEndSepPunct{\mcitedefaultmidpunct}
{\mcitedefaultendpunct}{\mcitedefaultseppunct}\relax
\EndOfBibitem
\bibitem[Scaffardi \latin{et~al.}(2004)Scaffardi, Pellegri, De~Sanctis, and
  Tocho]{scaffardi2004sizing}
Scaffardi,~L.~B.; Pellegri,~N.; De~Sanctis,~O.; Tocho,~J.~O. Sizing gold
  nanoparticles by optical extinction spectroscopy. \emph{Nanotechnology}
  \textbf{2004}, \emph{16}, 158\relax
\mciteBstWouldAddEndPuncttrue
\mciteSetBstMidEndSepPunct{\mcitedefaultmidpunct}
{\mcitedefaultendpunct}{\mcitedefaultseppunct}\relax
\EndOfBibitem
\bibitem[S{\"o}nnichsen \latin{et~al.}(2002)S{\"o}nnichsen, Franzl, Wilk, von
  Plessen, Feldmann, Wilson, and Mulvaney]{sonnichsen2002drastic}
S{\"o}nnichsen,~C.; Franzl,~T.; Wilk,~T.; von Plessen,~G.; Feldmann,~J.;
  Wilson,~O.; Mulvaney,~P. Drastic reduction of plasmon damping in gold
  nanorods. \emph{Physical review letters} \textbf{2002}, \emph{88},
  077402\relax
\mciteBstWouldAddEndPuncttrue
\mciteSetBstMidEndSepPunct{\mcitedefaultmidpunct}
{\mcitedefaultendpunct}{\mcitedefaultseppunct}\relax
\EndOfBibitem
\bibitem[Ngoc \latin{et~al.}(2015)Ngoc, Wiedemair, van~den Berg, and
  Carlen]{ngoc2015plasmon}
Ngoc,~L. L.~T.; Wiedemair,~J.; van~den Berg,~A.; Carlen,~E.~T.
  Plasmon-modulated photoluminescence from gold nanostructures and its
  dependence on plasmon resonance, excitation energy, and band structure.
  \emph{Optics express} \textbf{2015}, \emph{23}, 5547--5564\relax
\mciteBstWouldAddEndPuncttrue
\mciteSetBstMidEndSepPunct{\mcitedefaultmidpunct}
{\mcitedefaultendpunct}{\mcitedefaultseppunct}\relax
\EndOfBibitem
\bibitem[Brongersma \latin{et~al.}(2015)Brongersma, Halas, and
  Nordlander]{brongersma2015plasmon}
Brongersma,~M.~L.; Halas,~N.~J.; Nordlander,~P. Plasmon-induced hot carrier
  science and technology. \emph{Nature nanotechnology} \textbf{2015},
  \emph{10}, 25--34\relax
\mciteBstWouldAddEndPuncttrue
\mciteSetBstMidEndSepPunct{\mcitedefaultmidpunct}
{\mcitedefaultendpunct}{\mcitedefaultseppunct}\relax
\EndOfBibitem
\bibitem[Schoenlein \latin{et~al.}(1987)Schoenlein, Lin, Fujimoto, and
  Eesley]{schoenlein1987femtosecond}
Schoenlein,~R.; Lin,~W.; Fujimoto,~J.; Eesley,~G. Femtosecond studies of
  nonequilibrium electronic processes in metals. \emph{Physical Review Letters}
  \textbf{1987}, \emph{58}, 1680\relax
\mciteBstWouldAddEndPuncttrue
\mciteSetBstMidEndSepPunct{\mcitedefaultmidpunct}
{\mcitedefaultendpunct}{\mcitedefaultseppunct}\relax
\EndOfBibitem
\bibitem[Wang \latin{et~al.}(2019)Wang, Shi, Shen, Wang, Cronin, and
  Dawlaty]{wang2019ultrafast}
Wang,~Y.; Shi,~H.; Shen,~L.; Wang,~Y.; Cronin,~S.~B.; Dawlaty,~J.~M. Ultrafast
  dynamics of hot electrons in nanostructures: distinguishing the influence on
  interband and plasmon resonances. \emph{ACS Photonics} \textbf{2019},
  \emph{6}, 2295--2302\relax
\mciteBstWouldAddEndPuncttrue
\mciteSetBstMidEndSepPunct{\mcitedefaultmidpunct}
{\mcitedefaultendpunct}{\mcitedefaultseppunct}\relax
\EndOfBibitem
\bibitem[Logunov \latin{et~al.}(1997)Logunov, Ahmadi, El-Sayed, Khoury, and
  Whetten]{logunov1997electron}
Logunov,~S.; Ahmadi,~T.; El-Sayed,~M.; Khoury,~J.; Whetten,~R. Electron
  dynamics of passivated gold nanocrystals probed by subpicosecond transient
  absorption spectroscopy. \emph{The Journal of Physical Chemistry B}
  \textbf{1997}, \emph{101}, 3713--3719\relax
\mciteBstWouldAddEndPuncttrue
\mciteSetBstMidEndSepPunct{\mcitedefaultmidpunct}
{\mcitedefaultendpunct}{\mcitedefaultseppunct}\relax
\EndOfBibitem
\bibitem[Zhang \latin{et~al.}(2018)Zhang, Huang, Wang, Huang, He, and
  Wei]{zhang2018transient}
Zhang,~X.; Huang,~C.; Wang,~M.; Huang,~P.; He,~X.; Wei,~Z. Transient localized
  surface plasmon induced by femtosecond interband excitation in gold
  nanoparticles. \emph{Scientific reports} \textbf{2018}, \emph{8}, 1--7\relax
\mciteBstWouldAddEndPuncttrue
\mciteSetBstMidEndSepPunct{\mcitedefaultmidpunct}
{\mcitedefaultendpunct}{\mcitedefaultseppunct}\relax
\EndOfBibitem
\bibitem[Yeshchenko \latin{et~al.}(2013)Yeshchenko, Bondarchuk, Gurin, Dmitruk,
  and Kotko]{yeshchenko2013temperature}
Yeshchenko,~O.; Bondarchuk,~I.; Gurin,~V.; Dmitruk,~I.; Kotko,~A. Temperature
  dependence of the surface plasmon resonance in gold nanoparticles.
  \emph{Surface Science} \textbf{2013}, \emph{608}, 275--281\relax
\mciteBstWouldAddEndPuncttrue
\mciteSetBstMidEndSepPunct{\mcitedefaultmidpunct}
{\mcitedefaultendpunct}{\mcitedefaultseppunct}\relax
\EndOfBibitem
\bibitem[Liu \latin{et~al.}(2013)Liu, Ouyang, and Ye]{liu2013gold}
Liu,~L.; Ouyang,~S.; Ye,~J. Gold-nanorod-photosensitized titanium dioxide with
  wide-range visible-light harvesting based on localized surface plasmon
  resonance. \emph{Angewandte Chemie} \textbf{2013}, \emph{125},
  6821--6825\relax
\mciteBstWouldAddEndPuncttrue
\mciteSetBstMidEndSepPunct{\mcitedefaultmidpunct}
{\mcitedefaultendpunct}{\mcitedefaultseppunct}\relax
\EndOfBibitem
\bibitem[Clavero(2014)]{clavero2014plasmon}
Clavero,~C. Plasmon-induced hot-electron generation at nanoparticle/metal-oxide
  interfaces for photovoltaic and photocatalytic devices. \emph{Nature
  Photonics} \textbf{2014}, \emph{8}, 95--103\relax
\mciteBstWouldAddEndPuncttrue
\mciteSetBstMidEndSepPunct{\mcitedefaultmidpunct}
{\mcitedefaultendpunct}{\mcitedefaultseppunct}\relax
\EndOfBibitem
\bibitem[Hao \latin{et~al.}(2016)Hao, Guo, Pan, Chen, Jiao, Yang, and
  Guo]{hao2016visible}
Hao,~C.-H.; Guo,~X.-N.; Pan,~Y.-T.; Chen,~S.; Jiao,~Z.-F.; Yang,~H.; Guo,~X.-Y.
  Visible-light-driven selective photocatalytic hydrogenation of cinnamaldehyde
  over Au/SiC catalysts. \emph{Journal of the American Chemical Society}
  \textbf{2016}, \emph{138}, 9361--9364\relax
\mciteBstWouldAddEndPuncttrue
\mciteSetBstMidEndSepPunct{\mcitedefaultmidpunct}
{\mcitedefaultendpunct}{\mcitedefaultseppunct}\relax
\EndOfBibitem
\bibitem[Mukherjee \latin{et~al.}(2014)Mukherjee, Zhou, Goodman, Large,
  Ayala-Orozco, Zhang, Nordlander, and Halas]{mukherjee2014hot}
Mukherjee,~S.; Zhou,~L.; Goodman,~A.~M.; Large,~N.; Ayala-Orozco,~C.;
  Zhang,~Y.; Nordlander,~P.; Halas,~N.~J. Hot-electron-induced dissociation of
  H2 on gold nanoparticles supported on SiO2. \emph{Journal of the American
  Chemical Society} \textbf{2014}, \emph{136}, 64--67\relax
\mciteBstWouldAddEndPuncttrue
\mciteSetBstMidEndSepPunct{\mcitedefaultmidpunct}
{\mcitedefaultendpunct}{\mcitedefaultseppunct}\relax
\EndOfBibitem
\bibitem[Mukherjee \latin{et~al.}(2013)Mukherjee, Libisch, Large, Neumann,
  Brown, Cheng, Lassiter, Carter, Nordlander, and Halas]{mukherjee2013hot}
Mukherjee,~S.; Libisch,~F.; Large,~N.; Neumann,~O.; Brown,~L.~V.; Cheng,~J.;
  Lassiter,~J.~B.; Carter,~E.~A.; Nordlander,~P.; Halas,~N.~J. Hot electrons do
  the impossible: plasmon-induced dissociation of H2 on Au. \emph{Nano letters}
  \textbf{2013}, \emph{13}, 240--247\relax
\mciteBstWouldAddEndPuncttrue
\mciteSetBstMidEndSepPunct{\mcitedefaultmidpunct}
{\mcitedefaultendpunct}{\mcitedefaultseppunct}\relax
\EndOfBibitem
\bibitem[Johnson and Christy(1972)Johnson, and Christy]{johnson1972optical}
Johnson,~P.~B.; Christy,~R.-W. Optical constants of the noble metals.
  \emph{Physical review B} \textbf{1972}, \emph{6}, 4370\relax
\mciteBstWouldAddEndPuncttrue
\mciteSetBstMidEndSepPunct{\mcitedefaultmidpunct}
{\mcitedefaultendpunct}{\mcitedefaultseppunct}\relax
\EndOfBibitem
\bibitem[Malitson(1965)]{malitson1965interspecimen}
Malitson,~I. Interspecimen comparison of the refractive index of fused silica.
  \emph{Josa} \textbf{1965}, \emph{55}, 1205--1209\relax
\mciteBstWouldAddEndPuncttrue
\mciteSetBstMidEndSepPunct{\mcitedefaultmidpunct}
{\mcitedefaultendpunct}{\mcitedefaultseppunct}\relax
\EndOfBibitem
\end{mcitethebibliography}

\end{document}